\documentclass[11pt]{article}
\usepackage{setspace}
\usepackage[caption=false]{subfig} %% the caption=false ensures justified figure captions 
\usepackage{graphicx,caption}
\usepackage{authblk} 
\usepackage{indentfirst}
\usepackage[left=3cm,right=2cm,top=2.5cm,bottom=2cm]{geometry}
\usepackage{float}
\usepackage{array}
\usepackage{xcolor}
\usepackage{mathtools}
\usepackage{amsmath,amsfonts}
\usepackage{bbm}
\usepackage{hyperref}
\usepackage[capitalise,noabbrev]{cleveref}
\usepackage{todonotes}
\usepackage[utf8]{inputenc}
\usepackage{siunitx}
\usepackage{bm}
\usepackage{float}

\usetikzlibrary{patterns}
\newcommand*{\TitleFont}{%
      \usefont{\encodingdefault}{\rmdefault}{b}{n}%
      \fontsize{16}{20}%
     \selectfont}
\usepackage[percent]{overpic}

\makeatletter
\renewcommand{\maketitle}{\bgroup\setlength{\parindent}{0pt}
\begin{flushleft}
 \fontsize{16}{32}
 \textbf{\@title}
 
   \fontsize{12}{24}
  \@author
\end{flushleft}\egroup
}
\newfam\bboardfam
\font\tenbboard=msbm10  
 \font\sevenbboard=msbm7
   \font\fivebboard=msbm5 
\textfont\bboardfam=\tenbboard
\scriptfont\bboardfam=\sevenbboard
\scriptscriptfont\bboardfam=\fivebboard

\begin{document}
\author[1]{Zijie Qu}
\author[1]{Dominik Schildknecht}
\author[1]{Shahriar Shadkhoo}
\author[1]{Enrique Amaya}
\author[1]{Jialong Jiang}
\author[2]{Heun Jin Lee}
\author[1,2,3]{Rob Phillips}
\author[1]{Matt Thomson}
%\doublespace

%\title{\Large Persistent fluid flows defined by active matter boundaries}
\title{\TitleFont Persistent fluid flows defined by active matter boundaries}

\affil[1]{Division of Biology and Biological Engineering, California Institute of Technology, Pasadena, California, 91125, USA.}

\affil[2]{Department of Applied Physics, California Institute of Technology, Pasadena, California, 91125, USA.}

\affil[3]{Department of Physics, California Institute of Technology, Pasadena, California, 91125, USA.}

\maketitle

\subsection*{Abstract} 
\noindent
\textbf{Biological systems achieve precise control over ambient fluids through the self-or\-ga\-ni\-za\-tion of active protein structures including flagella, cilia, and cytoskeletal networks. In active structures individual proteins consume chemical energy to generate force and motion at molecular length scales. Self-organization of protein components enables the control and modulation of fluid flow fields on micron scales. The physical principles underlying the organization and control of 
active-matter driven fluid flows are poorly understood. Here, we apply an optically-controlled active-matter system composed of microtubule filaments and light-switchable kinesin motor proteins to analyze the emergence of persistent flow fields in a model active matter system. Using light, we form contractile microtubule networks of varying shape.  We analyze the fluid flow fields generated by a wide range of microtubule network geometries and explain the resulting flow fields within a unified theoretical framework. We specifically demonstrate that the geometry of microtubule flux at the boundary of contracting microtubule networks predicts the steady-state fluid flow fields across polygonal network geometries through finite-element simulations. Our work provides a foundation for programming microscopic fluid-flows with controllable active matter and could enable the engineering of versatile and dynamic microfluidic devices.}

\section*{Introduction} 
\indent
The precise control of fluids is essential for biological processes including motility and material transport across  cell, tissue and organismal scales of organization~\cite{allen1978cytoplasmic,behkam2007bacterial,berg1973bacteria,boon2013intercellular,pierce2008genetic,videler1993fish}. Biological systems generate and modulate fluid flows through the self-organization of ``active" protein structures like cilia and cytoskeletal networks that convert chemical energy into mechanical forces and motion to induce and modulate fluid flow. Active structures consume chemical energy and generate force at molecular scales but use self-organization of protein components to generate and control fluid flow fields that on micron length scales.  Cytoskeletal networks composed of filaments and motor proteins, for example, induce cytoplasmic flows or streams within cells that drive processes ranging from chloroplast transport in algae to the recycling of motility proteins during eukaryotic chemotaxis~\cite{allen1978cytoplasmic,keren2009intracellular,monteith2016mechanism} .  Understanding mechanisms of biological fluid control will enable the engineering of devices that use similar principles of self-organization to enable micron-scale control over fluids in technologies. 

The key conceptual challenge in cellular-scale biological fluid control is understanding how cells can apply molecular scale forces generated by active structures to modulate and organize fluid flows on mesoscopic length scales. The cell's reliance on self-organized, active structures like the cytoskeleton to control fluids complicates the analysis of biologically driven fluid flows. In a conventional system like a pipe or propeller, fluid flows emerge due to the macroscopic forces applied to a boundary conditions~\cite{batchelor2000introduction,wu2017transition,woodhouse2012spontaneous}.  However, in cellular flow control, force fields emerge from within the fluid itself,  and  force fields and boundary conditions emerge through molecular self-organization. Further, protein components both induce and respond to fluid flows, so that flows alter the distribution of force and material over time.  A fundamental challenge in analyzing and harnessing active-matter driven fluid flows is in elucidating how the interplay between molecular self-organization and material transport determines the architecture of active-matter driven fluid flows~\cite{monteith2016mechanism}. 

\begin{figure*}[p]
\centering
\includegraphics[scale=0.85]{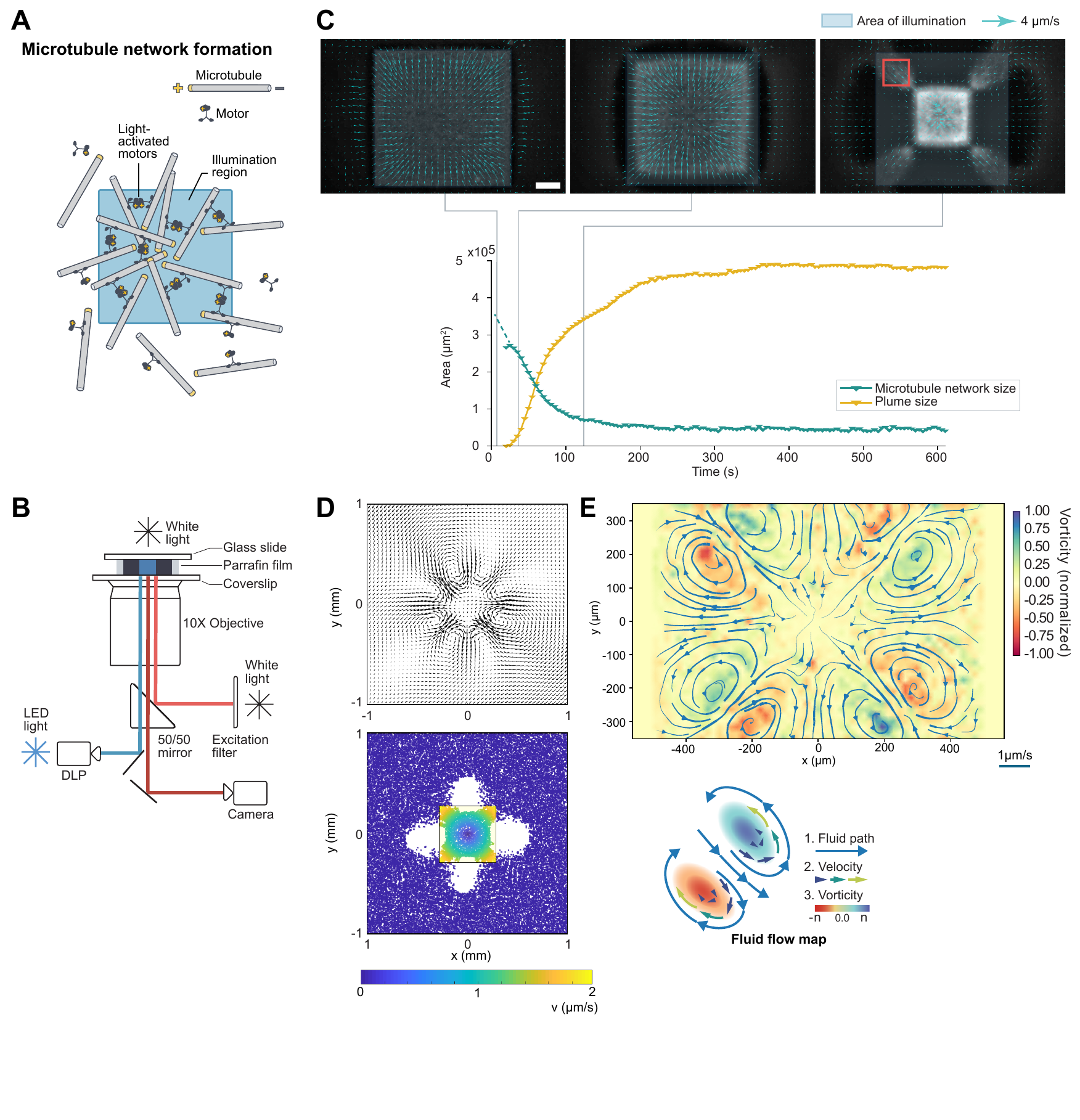}
\caption{\textbf{Constant illumination of the active matter system leads to microtubule network contraction and persistent fluid flows.} A. The active matter system used in this study consists of fluorescently labeled, stabilized microtubule filaments and kinesin motors that cross-link under illumination. B. The experimental setup: The microtubules are visualized with a fluorescence microscope; the cross-linking of the motors is controlled temporally and spatially by projecting light patterns using a digital light projector, the inert tracer particles are visualized with white light for measuring fluid flow. C. Three snapshots of the contracting network at $t =$~\SIlist{10;40;120}{\second} showing the contracting network with the emergence of arm-like microtubule bundles by projecting a \SI{600}{\micro\meter} square pattern; the cyan arrows demonstrate the steady-state fluid flow caused by the active force of the network; the scale bar is \SI{100}{\micro\meter}. The network and plume sizes are quantitatively measured with image analysis. D. A particle based simulation shows the emergence of vertex microtubule flux, each particle is modeled as a point force. E. Time-averaged flow field showing a persistent structure: four inflows along the diagonal of the projected square pattern, four outflows perpendicular to the edges and eight vortices. It is illustrated in the schematic that the vortices are located at both sides of the inflows.}
\label{fig:fig1}
\end{figure*}

Simplified active matter systems provide a reduced context in which to examine principles of self-organized fluid control in biological systems as well as a potential platform for building new technologies. Active matter systems composed of purified filament and motor proteins, specifically,  can generate non-equilibrium, self-organized structures including microtubules asters~\cite{hentrich2010microtubule,urrutia1991purified} and  contractile microtubule networks~\cite{ndlec1997self}. Further, active-matter systems~\cite{hitt1990microtubule} also generate spontaneous fluid flows~\cite{wu2017transition,sanchez2012spontaneous,guillamat2017taming,opathalage2019self}. However, spontaneous, active-matter driven fluid flows are typically disorganized and chaotic.  Artificial  boundaries can be applied to confine active systems experimentally and  organize fluid flows.  Recently, we developed an optically controlled active matter system~\cite{ross2019controlling} (\cref{fig:fig1}A) composed of microtubules and a engineered kinesin motor-proteins, where fluid flows can be generated and modulated with light. In the optically modulated system, organized fluid flows emerge through optically guided self-organization of microtubules and motors into networks that induce flow fields in ambient fluids. The optical experimental system provides a model in which to understand principles of fluid flow organization in active matter. While the geometry of light inputs into the system alters patterns of the fluid flow,   we lack a predictive understanding of how flow fields become spontaneously self-organized through active-matter dynamics. 

In this paper, we apply the light switchable active matter system to analyze emergence of self-organized flow fields within optically controlled active matter ~\cite{ross2019controlling}. We apply optical control to generate microtubule networks of different size and shape that induce persistent fluid flows. By using light to modulate the geometry of microtubule-motor networks,  we demonstrate that self-organized flow fields emerge through a dynamic feedback between microtubule network contraction and fluid driven mass transport. Fluid-network interactions generate persistent microtubule flux at the corners of contracting, polygonal microtubule newtworks.  The geometry of microtubule flux at the network boundary predicts the steady-state flow field for a wide range of polygonal microtubule network geometries. We develop a simple theoretical model based upon boundary point forces that predicts the qualitative and quantitative flow fields generated by polygonal microtubule network contraction. Our results reveal a dynamic mechanism of fluid flow generation in active microtubule networks and provides a modeling framework to predict the flow-field architecture from an active matter generated boundary condition.   This work lays the foundation for engineering microfluidic devices, where flow fields can be controlled in time and space with light.

\section*{Microtuble network contraction induces mass transport and persistent fluid flows }

To examine principles of active-matter driven fluid flows, we 
analyzed the emergence of the spontaneous fluid flow field within a recently developed active matter system in which fluid flows can be induced and modulated with light. The active matter system consists of stabilized microtubule filaments and kinesin motor proteins (\cref{fig:fig1}A; Methods). The kinesin motor protein has been engineered so that motors cross-link under illumination (illustrated in \cref{fig:fig1}A). Motor cross-linking induces interactions between motors on neighboring microtubules leading to microtubule network formation, contraction, and the generation of persistent fluid flows. In  previous work~\cite{ross2019controlling}, we described the emergence of active-matter driven fluid flows within the experimental system. However, the mechanism of flow generation and modulation has remained poorly understood. 

We developed an experimental platform~\cite{ross2019controlling} that allows us to optically induce and track both microtubules and flow-field dynamics.  Our experimental system (\cref{fig:fig1}B) uses a conventional fluorescence microscope and a digital light projector to control sample illumination. We track microtubule dynamics through a fluorescence microtubule dye and, simultaneously, introduce micron sized tracer particles to quantify the flow-field architecture through particle tracking. The experimental system allows us to induce microtubule networks of differing size and shape and to track the dynamics of the active matter-fluid system as the fluid flows emerge. 

\begin{figure}[p]
\centering
\includegraphics[width=11cm]{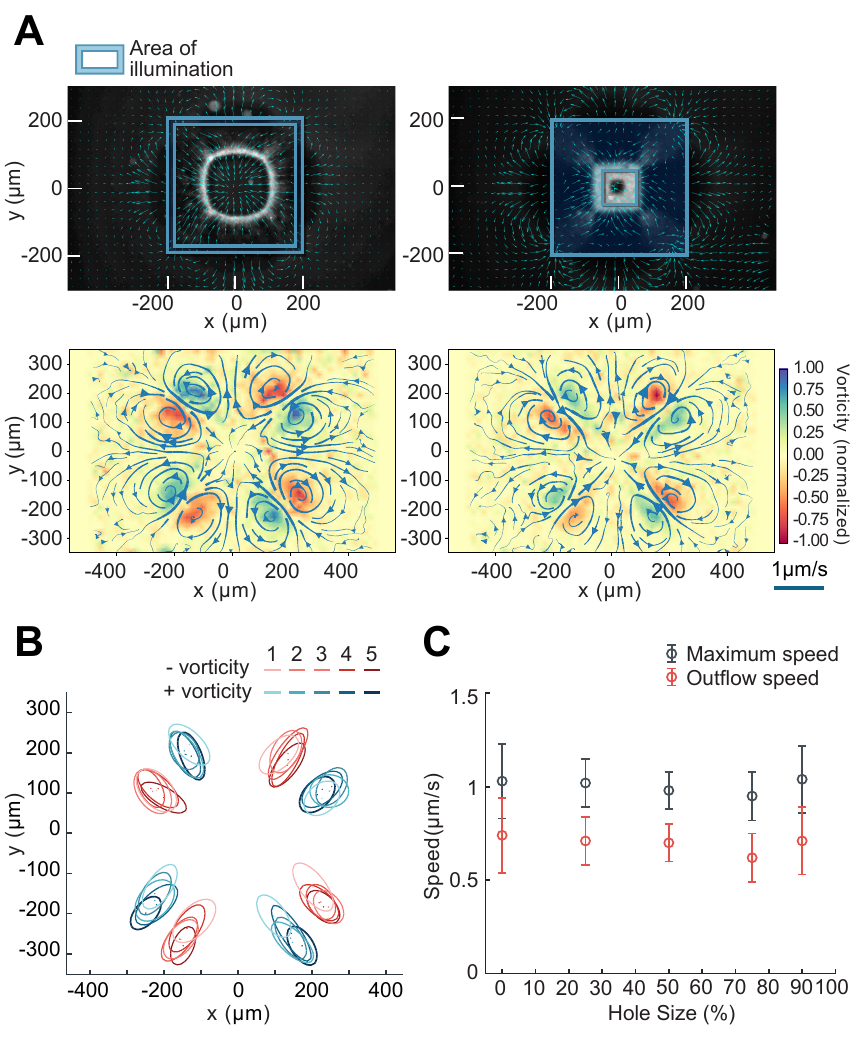}
\caption{\textbf{The fluid flow for square illumination patterns is not affected by introducing a cavity at the network center.} A. The projected light patterns are hollow squares (\SI{400}{\micro\meter} in length) with different cavity sizes; left: the cavity spans 90\% of the square's size, right: the cavity spans 25\% of the square's size; time-averaged flow fields on the bottom (more cases in SI); the initial contracting network demonstrates different behavior, yet the microtubule fluxes are the same as demonstrated by the PIV result; the resulting flow fields are persistent and show similar structures -- four inflow along the diagonal directions and outflow from the edges, plus eight vortices. B. The vorticity peaks and valleys from different hollow square experiments overlap. C. The maximum inflow and outflow velocities are similar across all experiments, the data are averaged over five individual experiments, each experiments have four maximum inflow and outflow values; the error bar shows the standard deviation.}\label{fig:fig2}
\end{figure} 

We first analyzed the fluid dynamics induced by optical activation of a  \SI{600}{\micro\meter} square region (\cref{fig:fig1}C). We found that persistent flows are generated through a dynamic process that is initiated by microtubule network contraction and that leads to a persistent influx of microtubules along the diagonals of the square network (\cref{fig:fig1}C). Specifically,  light activation induces the formation of a microtubule network that spontaneously contracts to its center of mass. As the microtubule network contracts, it induces flows within the bulk microtubules (\cref{fig:fig1}C, cyan arrows) outside of the light-activated region as quantified by Particle Image Velocimetry (PIV). The background flows lead to an influx of microtubules along the corners of the contracting microtubule network. Persistent microtubule flux at the network corners is maintained for more than $30$ minutes due to a persistent influx of microtubules from the background solution into the light-activated region. 

We observed similar microtubule flux in simulations of contracting square microtubule networks (Methods). In these simulations, we treat each position within a contracting microtubule network as a point force whose amplitude is set by the contractile velocity of the network.  Each point force acts as a source for fluid flow due to Stokesian fluid dynamics~\cite{mogilner2018intracellular} (\cref{fig:fig1}D). In these simulations microtubule network contractions generate a fluid flow field that naturally lead to inflows of particles at the network corners. Microtubule induced flows have been observed and modeled in other biological contexts and motile microtubules have been modeled as point forces that act locally to induce fluid flow~\cite{mogilner2018intracellular,needleman2019stormy}

Experimentally, the emergence of persistent microtubule flux at the corners of the contracting network coincides with the emergence of steady-state fluid flows (\cref{fig:fig1}E) within the system as measured by particle tracking. Specifically, persistent fluid flows emerge within the system with inflows of fluid along the corners of the square and fluid outflows along the edges of the activation region. In addition to the fluid inflows and outflows, the fluid flow field contains four pairs of approximately \SI{100}{\micro\meter} in radius centers of vorticity  oriented along each arm of the microtubule structure.  Broadly, fluid inflows align with the persistent influx of microtubules along the microtubule network corners.  The observation is consistent with a model where microtubule inflows along the corners of the network act as force sources that generate the fluid flows.  

Thus, optical induction of a square microtubule network leads to a persistent fluid flow field, whose architecture coincides with active streams of microtubules transported along the diagonals of the microtubule network. In the persistent configuration of the system, high density regions of microtubules emerge on the corners of the network, where persistent microtubule influx occurs from the bulk into the active region of the network. Coincident with these microtubule flows are persistent fluid flows in the background fluid.

%(ref) and the density of force generated by a active microtubule fluid is proportional to the velocity of the microtubules times and density of material.

%The dynamics of microtubule network contraction and fluid induced mass transport leads to a persistent influx of microtubules along the diagonal of the microtubule network. The persistent influx continues for X minutes.Microtubule density becomes concentrated at the corners of the contracting network and a persistent in-flows of material, at um per second, occurs on the network boundary so that a persistent flux of active material generates flows of microtubules into the center of the network.   During this time the velocity of microtubules along the diagonal is approximately X um/sec, and the density of microtubules is approximately XX10mg/ml. As the network contracts the microtubule influx becomes persistent generating a source of force within the system. 

%Microtubules actively moving through a fluid act as force dipoles (ref) and the density of force generated by a active microtubule fluid is proportional to the velocity of the microtubules times and density of material. In our system, the velocity density which is proportional to the force density is maximum along the diagonal of the network. 

%Option: Contraction generated mass transport induces microtubule vertex flux in simulation

% Vertex flux model predicts fluid flow fields 

\section*{Microtubule network boundary geometry determines organization of flow field}
 
We found that the geometry of the microtubule network boundary alone is sufficient to determine the architecture of the microtubule streams and fluid flow fields for square microtubule networks. To analyze the role of the microtubule flux along the boundary of the microtubule network in generating and organizing fluid flows, we created a series of \SI{400}{\micro\meter} microtubule networks with a square boundary where we introduced square cavities into the center of the network (\cref{fig:fig2}A). In total, we created four different networks with cavities ranging from \SI{100}{\micro\meter} to \SI{360}{\micro\meter} in length. By altering the distribution of active material and the network topology,  the hollow networks allowed us to examine the relative contribution of the network boundary and interior to the generation of the fluid flow field. 

The hollow square networks execute a similar process of microtubule network formation, network contraction, and boundary linking, leading to the persistent microtubule flux at network corners. For the hollow shapes, the contractile network itself adopts the geometry of the light pattern, so that network itself contains a square hole throughout the network contraction (\cref{fig:fig2}A).  In each case, the contracting network generates a diagonal distribution of microtuble flux at the corners of the illumination region (\cref{fig:fig2}A). The microtubule distribution observed for the hollow networks is qualitatively similar to the material distribution generated by contraction of the intact square network. Like the intact square network in \cref{fig:fig1}C, the hollow networks generate streams of microtubules along the network corners, an accumulation of microtubule density, and 
a persistent microtubule flux. 

\begin{figure*}[p]
\centering
\includegraphics[scale =.8]{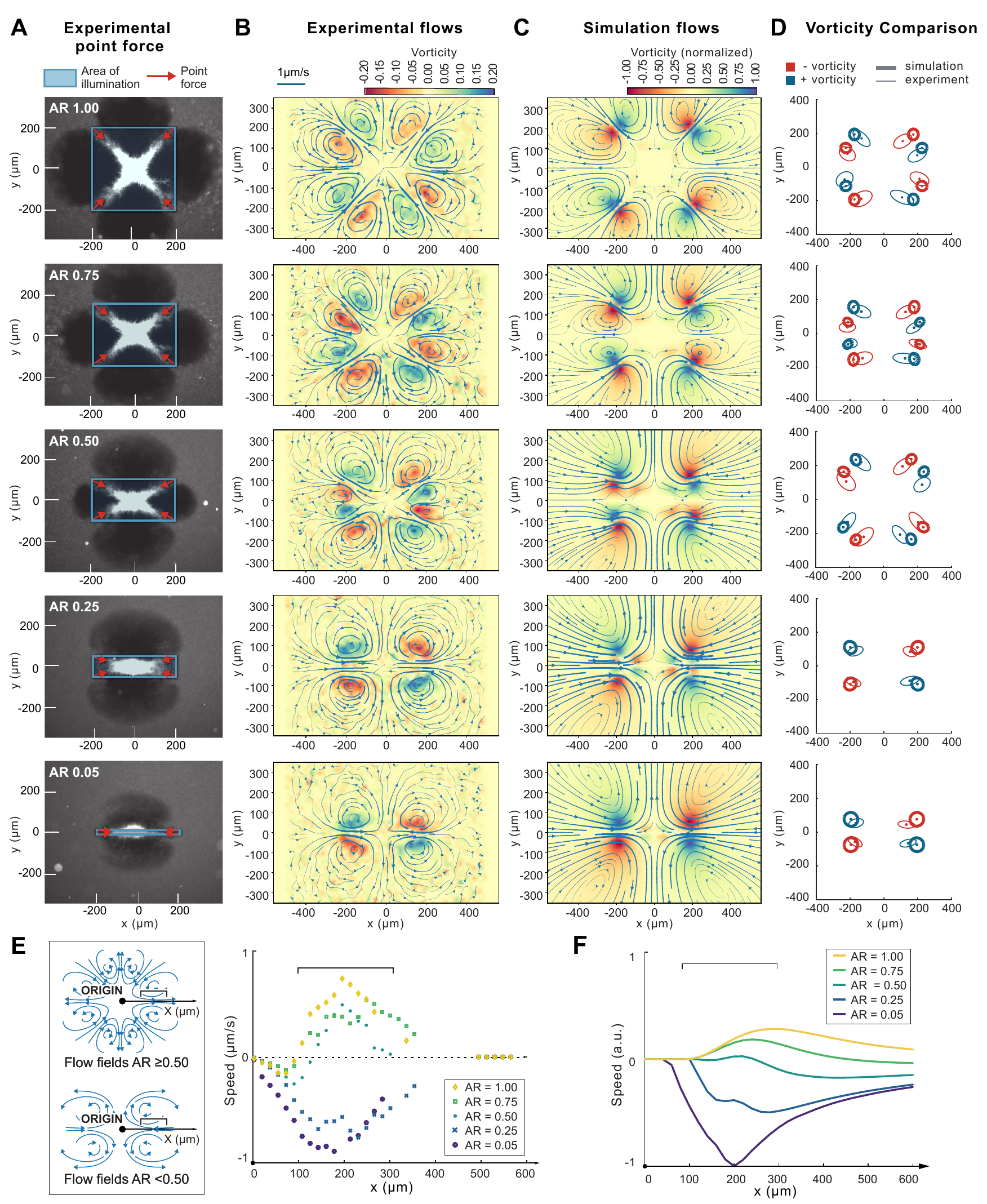}
\caption{\textbf{The fluid flow fields agree qualitatively and quantitatively with the predictions from the boundary force model.} A. Snapshots from the experiments showing the projected light patterns, the length of the rectangular pattern is \SI{400}{\micro\meter} and the aspect ratios are $1,~\frac{3}{4},~\frac{1}{2},~\frac{1}{4},$ and $\frac{1}{20}$ from top to the bottom; the red arrows indicate the locations of the point force used in the simulation. B. Resultant time-averaged flow fields. C. Simulated flow fields with a finite element method using point forces as source term. D. The peaks and valleys of the vorticity from experimental and simulation results show a good agreement. E. Left: A schematic showing the location where the velocity is measured and compared; Right: The horizontal component of the fluid velocity measured experimentally along $x$-axis, each data point is averaged over ten individual experiments. F. The horizontal component of the fluid velocity measured from simulation result along $x$-axis, similar trend is observed compared to the experimental results in \cref{fig:fig3}E.
\label{fig:fig3}}
\end{figure*}

Quite surprisingly, the hollow squares also generate persistent fluid flow fields (\cref{fig:fig2}A, lower plots) that are nearly identical to the intact square networks in terms of their qualitative and quantitative features. Like the filled square network, hollow-network-induced flow fields exhibit diagonal inflows, edge outflows, and a series of vortex pairs.  The vortex pairs, specifically, across all networks were quantitatively comparable in their size and location (\cref{fig:fig2}B). We quantified the position of vortices by fitting the vorticity ($\nabla \times v$) with a Gaussian mixture model and extracting the centroid and shape of each vorticity peak or valley.  The extracted vorticity peaks across the five different networks had centroids centers lying within \SI{10}{\micro\meter} of one another (\cref{fig:fig2}B).  Further, the maximum of the fluid velocity was approximately \SI{1}{\micro\meter\per\second} along the inflow and \SI{0.7}{\micro\meter\per\second} along the outflow (\cref{fig:fig2}C), and the maximal and minimal fluids speeds agree across the five networks within the variation of measured for replicates of individual networks. 

The quantitative similarity between the flow fields induced by the hollow networks suggested that the the geometry of the microtubule network boundary is sufficient to determine the architecture of the induced flow field. Specifically, the distribution of microtubule density and flux at the corners of the applied light pattern was similar for the four cavity networks, suggesting that the microtubule flux at the corners of the  contracting  networks plays a central role in determining the architecture of the generated fluid flows. 

\section*{Boundary force model predicts flow field geometry}
Based upon our analysis of the flow fields induced by the hollow square networks, we developed a simple mathematical model to predict the architecture of fluid flow fields generated by contracting microtubule networks. In this model, the microtubule influx at the corners of the network generates force that that is transferred to the fluid to induce flow. Our model is based upon the following observations. First, microtubule network contraction leads to diagonal inflows of activated microtubules at the corners of the contracting microtubule network. Second, networks with a square boundary but distinctly different internal structure induce nearly identical patterns of fluid flow. Third, moving microtubules are known to act on their ambient fluid as local point sources of force. Therefore, we propose that the force generated by active microtubule flux along the corners of the contracting network determines the architecture of the induced flow field. 

In our model, the fluid is considered in the incompressible Stokesian limit (valid at low Reynolds numbers) so that the fluid velocity $\vec u$ obeys the Stokes equation  
\begin{align}
    \mu \nabla^2\vec u-\nabla p+\vec f&=0,  &\nabla \cdot \vec u &=0,\label{eq:stokes-eq}%
\end{align}
where $\mu$ is the fluid viscosity, $p$ is the pressure, and $\vec f$ is the force field exerted by the active matter. The hypothesis is that the geometry of the corners determines the flow. In this model, it means $\vec f$ to be non-vanishing only in the corners. Therefore, $\vec f$ is a sum of four point-forces directed towards the center of the shape for a rectangular pattern illumination, due to symmetry. These properties determine \cref{eq:stokes-eq} up to a trivial rescaling of all quantities, since \cref{eq:stokes-eq} is linear.

Due to the linearity of \cref{eq:stokes-eq}, the fundamental solution could be used to obtain the flow field for free boundary conditions~\cite{Chwang1975}. However, incorporating more general boundary conditions in this solution method can be challenging~\cite{Blake1971}. In this particular case, two sets of boundary conditions have to be imposed. The first set models the no-slip condition at the limits of our fluid flow cell, while the second set describes the inhibition of fluid flow due to the jamming of microtubules. This latter boundary condition is achieved by imposing an area with no-slip at its boundary the center of the activated area, with the same shape as the activation area, but with half of the (linear) dimensions. Since these boundary conditions are difficult to treat with the fundamental solution, finite-element (FEM) simulations were used, based on FeniCS~\cite{AlnaesBlechta2015a}. It should be noted that due to the limitations of FEM, point forces are spread out over finite kernel. A more detailed discussion of the implementation can be found in the appendix.  %\cref{}.

\section*{Boundary force model captures aspect ratio induced flow-field transitions for rectangular networks}
For square networks we found that the model predicted a flow-field with fluid in-flows along the network diagonal and outflows along the edges leading to pairs of vortices along the corners of the networks (\cref{fig:fig3}A - D, top row). In general, there is a striking resemblance between the qualitative features of experiment and theory can be observed. 

We, therefore, generalized our model to predict the flow generated by rectangular networks with a range of different aspect ratios. For rectangular networks, the model predicted the correct number and position of fluid vortices (\cref{fig:fig3}C).  Interestingly, in the rectangular geometries, we experimentally observe a crossover from in-flow to outflow along the long axis of the rectangle for increasing aspect ratio. %The cross-over is correctly described by the corner-force model and predicted to occur between an aspect ratio of $\frac{1}{4}$ and $\frac{1}{2}$ consistent with experimental data. 
Moreover, the emergence of new vertices left and right of the activated area can be well observed in the streamlines for both the experiment and theory. It is even more apparent in the vorticity, shown in the background of the figures, demonstrating the resemblance between theory and experiment.
%The results of these FEM simulations are compared to the experimental data in \cref{fig:fig3}(a) (third and second column, respectively).

%% paragraph: Discuss results II: quantitative: compare vorticities
The vorticity can be further analyzed by comparing the experiment to the theory directly. For this purpose, the vorticity from theory and experiment is fitted by a mixture of Gaussians. Their centers and their extent are then compared in the \cref{fig:fig3}D, where a significant overlap can be observed between experimental and simulated centers of vorticity. Direct comparison between the centers of the corresponding Gaussians shows that the average distance between experimental data and simulation is approximately \SI{46}{\micro\meter}, which is much smaller than the typical extent of the activated region of \SI{400}{\micro\meter}.

%\todo{maybe rather compare with the typical diameter of ellipses?}

%% paragraph: Discuss results III: Comparison u_x(x) (Figs 3b/c)
Furthermore, while the number of vortex cores and their location is our primary method of comparing the simulation results with the experiment, the fluid velocities can also be compared directly. An interesting place to compare is along the long-axis of the rectangle, because the fluid velocity exhibits a (distance-dependent) crossover behavior from inflow to outflow for increasing aspect ratio. Qualitatively, we show that this crossover is revealed for both experiment and simulation in \cref{fig:fig3}E and F, respectively. However, we are cautious in the direct comparison, because the details of the crossover, such as the location of the inversion point at a fixed aspect ratio, depend on details of the simulation, such as the size of the inner jamming region, and are difficult to locate precisely in the experiment.
%Finally, the models capture the dependence of inflow versus outflow as a function of distance away from the illuminated region. Of particular interest for such comparisons are high-symmetry directions of the rectangle. Here, we focus on the high-symmetry line along the long axis of the rectangle, because this particular crossover is apparent in \cref{fig:fig3}A2, A3. Along this line, the $y$-component of the fluid velocity has to vanish, due to symmetry. Hence, it is sufficient to compare the $x$-component of the fluid velocity as a function of distance from the center of the rectangle in \cref{fig:fig3}B and C for the experimental data and the theoretical model, respectively. As already observed from the streamlines in \cref{fig:fig3}A2, A3, the inflow [$u_x(x)<0$] to outflow [$u_x(x)>0$] crossover behavior with aspect ratio is well observed in \cref{fig:fig3}B and C as well. Furthermore, the additional crossover behavior dependence on the distance from the activated area for intermediate aspect ratios can be seen for both the experiment and the theory, further confirming that the corner-force model is sufficient to describe the experimental data. 

\begin{figure*}[h!]
\centering
    \includegraphics[width=15cm]{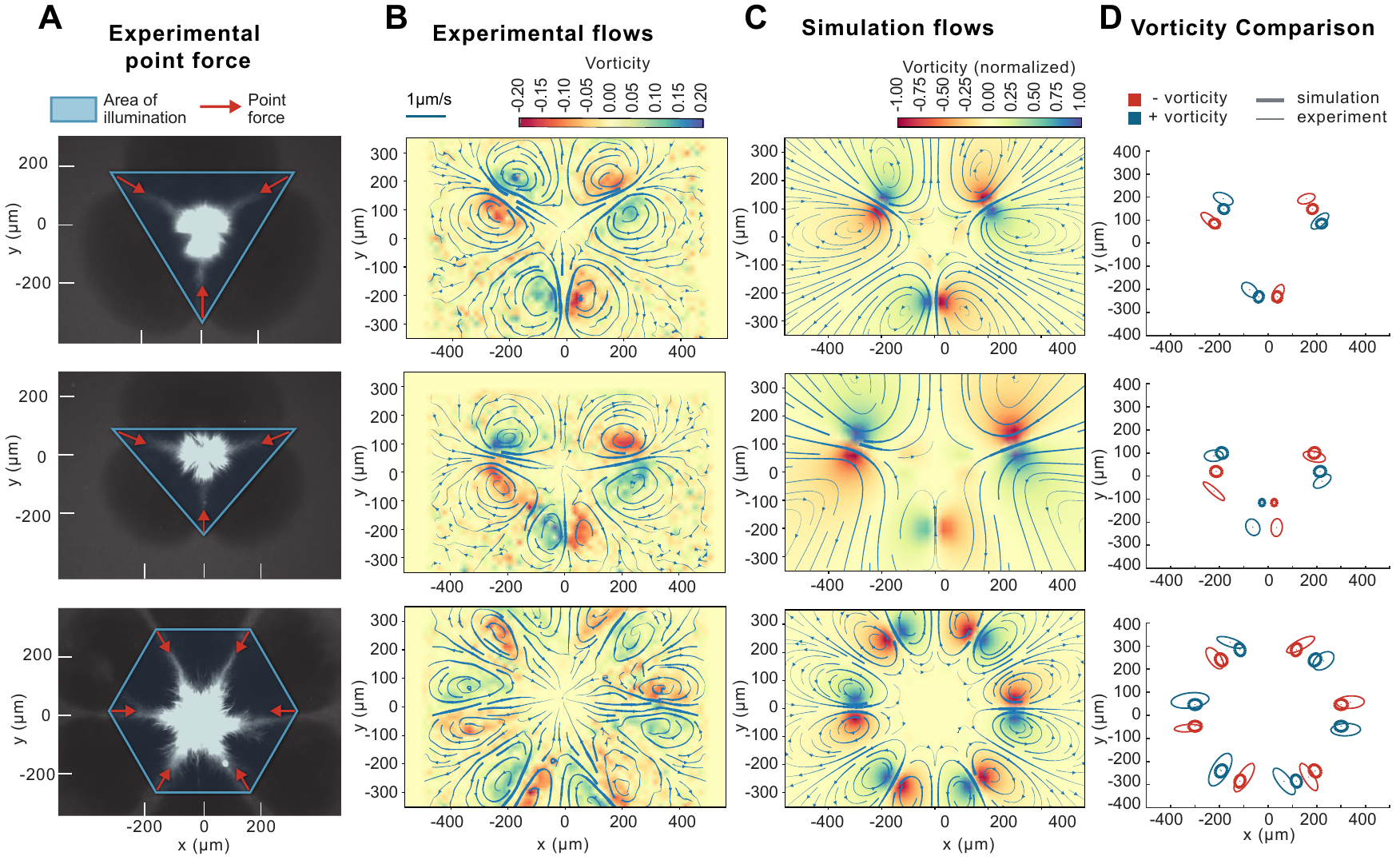}
\caption{\textbf{The simulations can be successfully generalized to other polygonal shapes, such as triangles and hexagons.} A. Snapshot images from the experiment showing the projected pattern (illustrated by light blue), an equal-angle triangle (\SI{600}{\micro\meter} in length), an isosceles triangle (\SI{600}{\micro\meter} in width, \SI{300}{\micro\meter} in height), a hexagon (\SI{600}{\micro\meter} from top to the bottom), the red arrows indicate the locations of the point force used in the simulation; B. Time-averaged flow fields measured experimentally; C. Simulated flow fields using point forces as source terms; D. 
The peaks and valleys of the vorticity from experimental and simulation results show a good agreement.}\label{fig:fig4}
\end{figure*}

\section*{The boundary force description predicts flow field induced by polygonal networks }
%% paragraph: Wish to generalize
%%     Rectangles are neat, but to have a truly broad toolset generalization required
%%    Validate the point force description by other shapes -> consider polygons
While the rectangular activation area discussed in the previous section provides a useful test case for the model, we are interested in developing a more general framework. Therefore, we generalize our framework to predict the flow fields generated by a range of polygonal shapes. Further, we introduce a point force onto the vertices of the polygonal light pattern. In each case, we found that the corner-force description was sufficient to determine the number and location of fluid vortices as well as the orientation of fluid flows. 

%% paragraph: regular polygons
%%    No issue for regular polygons (equal angles and sides) since all corners are treated the same due to symmetry.
%%     Indeed, for triangle and hexagon, the simulations indeed trivially agree (as observed in Fig 4
First, regular triangles and hexagons are considered. Due to symmetry, forces are in the corners, pointing towards the center, and each force amplitude is the same. Similar to the previous section, the results are compared in \cref{fig:fig4}B and C: The experimental and theoretical fluid flows are depicted, respectively. The first row depicts the regular triangle case, where the last row depicts the hexagon case. Again, an excellent qualitative agreement between theory and experiment can be observed, further demonstrated in \cref{fig:fig4}D, where the Gaussian mixture model comparison is shown. The locations of the vorticity peaks and valleys overlap well between the experimental observation and the simulation.
Hence, these comparisons demonstrate that our model can generalize to regular polygons.

%% paragraph: irregular polygon
%%    Consider a final example: isosceles triangle
%%     Inequality of corners, how do the forces need to scale? Answer -> Force proportional to the distance from center (experimentally justified by Tyler, theoretically by Shai/Jialong?)
%%     Using this, flow patterns again agree, see Fig 4
Finally, a non-regular polygon was also considered, through an isosceles triangle. Because the polygon is no longer regular, there is no longer a symmetry reason for all forces to be equally strong. While we still assume that the forces point towards the center of the activated area, we hypothesize that their amplitudes are scaling linearly with distance to the center. In other words, corners further away from the center induce a stronger fluid flow, which is motivated by experimental results in Ref.~\cite{ross2019controlling} and by a recent theoretical description~\cite{zijie2020self-affine}. 
The results of the model and the experiment are compared in the middle row of \cref{fig:fig4}B, C and D: Once again, a good agreement can be observed, therefore providing compelling evidence that our description can be applied to non-regular polygons.

\section*{Discussion}
In this work, we analyze the emergence of self-organized fluid flows in an optically controlled microtubule-motor active matter system \emph{in vitro} where the geometry of the active matter induced flow field can be modulated by an applied light pattern. We demonstrate that active-matter fluid interactions lead to self-organized flux of microtubules along the corners of polygonal microtubule networks. By modeling these fluxes as point forces, we predict the architecture of persistent flow fields generated by a range of polygonal  network geometries. Our model shows that the distribution of forces on the boundary of an active structure is sufficient to predict the flow field generated by the active matter system.   The corner-force field itself emerges through a self-organization process driven by microtubule network contraction and mass transport. Therefore, our model reveals how self-organized microtubule flux can control the geometry of active matter driven fluid flows.

In our model, microtubule flux along the boundary of the active matter system plays an essential role in determining the induced flow field. The persistent, steady-state flow structure observed in this work reveals the importance of force and material localization in active-matter-driven fluid flows. Our experimental system spontaneously generates organized flow fields that resemble extensional fluid flow fields. 

Previous work on active matter driven fluid flows has observed that  active nematics composed of microtubules and motor proteins generate spontaneous but chaotic fluid flows fields~\cite{sanchez2012spontaneous,guillamat2017taming}. Furthermore, in these systems confinement of the nematic within passive fabricated boundary leads to organization of chaotic flows~\cite{allen1978cytoplasmic,opathalage2019self}. In our system, self-organization of microtubule density within the active matter-fluid system itself generates organized flow fields. The boundary forces used in our predictive model emerge experimentally through the self-organization of microtubule `arms' at the corners of our applied light pattern. Therefore, our work reveals a mechanism of self-organized flow modulation and demonstrates how organized flows can emerge in active matter without passive, fabricated boundaries. Geometric control of active matter through self-organization is commonly observed in nature as cells and microorganisms use both physical and chemical mechanisms to control the shape of active matter networks to modulate  fluid flows~\cite{alim2013random,mogilner2018intracellular}.

We explored a limited set of flow patterns generated by convex, polygonal network geometries. It will be important, in future work, to explore the range of fluid fields that can be generated by exploring a wider range of geometries, considering possible history dependence and combining flow fields generated by different illumination patterns with super-position. Further, the organized flow field is a potential platform for building micro-machines. For example, both Sokolov \emph{et al.}~\cite{sokolov2010swimming} and Vizsnyiczai \emph{et al.}~\cite{vizsnyiczai2017light} showed that microscopic gears can be driven by bacterial suspensions to collect mechanical energy, a similar device can be combined with the microtubule-kinesin system and driven by the fluid flow as an active switch. The persistent flow observed in this active matter system sheds light on various flow control strategies on microscopic scale and can be used to resolve many emerging challenges in the design of microfluidic devices.

\section*{Summary Methods}
The active matter system consists of stabilized microtubules, kinesin motors (constructed with light-induced hetero-dimer system) and necessary energy mix. All ingredients and buffer preparation protocols are documented in a previous paper from our group \cite{ross2019controlling} and we follow the exact same procedure in this study using the aster assay. Inert particles with a \SI{1}{\micro\meter} diameter are suspended in the system for flow field measurement purposes. The sample chambers are made by sandwiching pre-cut Parafilm M by coated slides and coverslips \cite{ross2019controlling,Lau_acrylamidecoat2009}. The measured depth of the chamber is approximately \SI{70}{\micro\meter}.

The experiment is conducted on a conventional microscope (Nikon TE2000) with ten-fold magnification. We customize the system by adding a programmable digital light projector (EKB Technologies DLP LightCrafter E4500 MKII Fiber Couple), which is used to activate the kinesin motors with projected polygonal patterns. The DLP chip is illuminated by the \SI{470}{\nano\meter} LED (ThorLabs M470L3). Fluorescently labeled microtubules are illuminated by \SI{660}{\nano\meter} light and imaged with a digital camera (FliR BFLY-U3-23S6M-C). The system is controlled with Micro-Manager on a PC.

The microtubule flow field is estimated using particle image velocimetry (PIV) by analyzing the images of  fluorescently labeled microtubules (\SI{660}{\nano\meter} light). The particle image velocimetry is performed by a custom code written in Matlab~\cite{MATLAB2019}  to measure the velocity field of the microtubule contraction dynamics (within the illumination region) and microtubule movements alongside the fluid flow (outside the illumination region). The microtubule velocity field is determined through local maximization of the correlation between the warped and the target images.  

For the simulation in~\cref{fig:fig1}D, active particles are modeled as point forces that generate 2D Stokeslets~\cite{Spagnolie2012}. The force vectors point towards the center of the contracting network with a magnitude that is a linear function of the particle's distance from the network center.

The flow field is measured by first comparing the position of individual inert particle from frame to frame~\cite{adamczyk19882} for the particle displacement with a nearest-neighbor algorithm (\num{15} pixels threshold)~\cite{schmidt1996imaging}. Then the displacement vector field, which is sparse due to the randomness of the particle locations, is grouped within a $30 \times 30$ pixels consecutive window and averaged. The velocity field is also calculated by dividing the displacement field by \SI{5}{\second}, the time interval between frames. Finally, the velocity vector field at each time step of the experiment is summarized and averaged to get the time-averaged velocity field. 

%The fluid flow field is measured by tracking individual inert particle from frame to frame~\cite{adamczyk19882}. The inert particles are visualized using bright-field microscopy. Particles are identified and tracked using custom Matlab code. 

The solution to \cref{eq:stokes-eq} with the boundary conditions described in the text was found by employing finite-element methods. In particular, the FeniCS library was used~\cite{AlnaesBlechta2015a}. Our code can be found on GitHub~\footnote{\url{https://github.com/domischi/StokesFEM}}. No-slip boundary conditions were applied at the upper and lower edge of the simulation cell, as well as in a central region with half the linear size of the activation region. The point forces were spread out over a small region, with one tenth of the size of the activated region. The solution approach further used Taylor-Hoods basis functions and a quadratic meshing. Additional details can be found in the appendix.

\section*{Acknowledgement}
We acknowledge funding from the Donna and Benjamin M. Rosen Bioengineering Center, Foundational Questions Institute and Fetzer Franklin Fund through FQXi 1816 , Packard foundation, and Heritage Medical Research institute. Inna-Marie Strazhnik for preparation of figures and illustrations.

\section*{Appendix}

\subsection*{Flow cell}
The flow cell is made by sandwiching a paraffin film (Parafilm M) between a glass slide and a coverslip, both are treated with a hydrophilic acrylamide coating~\cite{lau2009condensation, ross2019controlling}. The paraffin film is heated at \SI{65}{\celsius} on top of the glass slide for 10$s$, then the coverslip is put above the paraffin film and generally pressed for 15$s$ until properly sealed. The final size of the flow cell is \SI{50}{\milli\meter} in length, \SI{3}{\milli\meter} in width and approximately \SI{70}{\micro\meter} in depth.

\subsection*{Measuring the flow field with particle tracking}
The flow field is measured by tracking individual particle from frame to frame~\cite{adamczyk19882}. The inert particles are visualized using bright-field microscopy. The images are pre-processed with a Gaussian filter in Matlab (MathWorks)  with standard deviation of 1 to reduce the noise level, then a binary filter to distinguish from the backgrounds. The centroid of each particle is measured by the \emph{regionprops} function in Matlab. The particle pairs are determined with a nearest-neighbor algorithm~\cite{schmidt1996imaging} with a \SI{8.8}{\micro\meter} distance threshold. The resultant distance vectors are then grouped and averaged within a \SI{17.6}{\micro\meter}. Finally, the vector fields are averaged temporally over the whole experimental process.

\subsection*{Particle image velocimetry analysis of microtubule flow field}
The particle image velocimetry is performed by a custom code written in MATLAB~\cite{MATLAB2019}. The microtubule velocity field is determined through local maximization of the correlation between the warped and the target images. Each image is first tiled by boxes of the size of the approximate correlation lengths of the velocity field ($\ell$). The boxes search their vicinity to find the direction and magnitude of the translation vector that maximizes the correlation with the target image. The velocity field is eventually smoothed out by applying a Gaussian kernel of the same width as the box size. The stability of the algorithm against box sizes are checked by varying them over the range of $[\ell/2\, ,\,2\ell]$.

\subsection*{Image Segmentation, microtubule network size analysis}
Traditional Computer Vision techniques available in Python~3.7~\cite{2020SciPy-NMeth, van2014scikit} were applied to segment microtubule network regions in every frame of the movie. In the pre-processing steps, we down-scaled original images by a factor of $\frac{1}{4}$ to make computation faster. Then we proceeded to correct the unevenness in illumination by computing a "strong" Gaussian blur and subtracting the resulting filter from every frame. We reduced the noise further by filtering the images using a \emph{scipy} median filter.
 
To segment the microtubule networks from the background we applied \emph{skimage} implementation of unsupervised thresholding algorithms; Otsu and triangle algorithms~\cite{otsu1979threshold, zack1977automatic}. The resulting microtubule network binary masks were assessed qualitatively, and we performed a series of binary morphological operations (erosion, dilation, opening, and closing), to remove unwanted small objects and to fill holes inside the area of the microtubule network.

%Plume binary masks were obtained by using state of the art supervised Machine Learning. Specifically, semantic segmentation was carried out using a Convolutional Neural Network (CNN) provided by \emph{DeepCell}; a python deep learning library for general purpose biological image segmentation~\cite{van2016deep}. 

%All deep learning steps were performed using downscaled images (factor 1/4). The DeepCell CNN was trained to classify individual pixels as belonging to one of three classes: aster, plume, or background. To train the model we used 1007 manually labeled images. This training dataset corresponds to five full-length active matter movies and each movie was produced using a unique light pattern. 

%After training, we ran predictions for classifying the pixels in the data of interest. We only post-processed the probability map corresponding to plume values.  We manually thresholded the probability maps to produce the final plume binary masks. Since there was some overlapping between the plume masks and the aster masks, we trimmed the plume masks by making zero all pixels inside the plume masks that belonged to aster regions in the aster mask. 

To quantify the area corresponding to either microtubule network or plume, we up-scaled all masks to match the original raw image size. After that, we used \emph{numpy} to count all non-zero pixels in each mask and multiply the resulting count by the interpixel distance squared: $(\SI{0.586}{\micro\meter})^2$. 

\subsection*{Theoretical Model and the corresponding Finite-element Simulation}
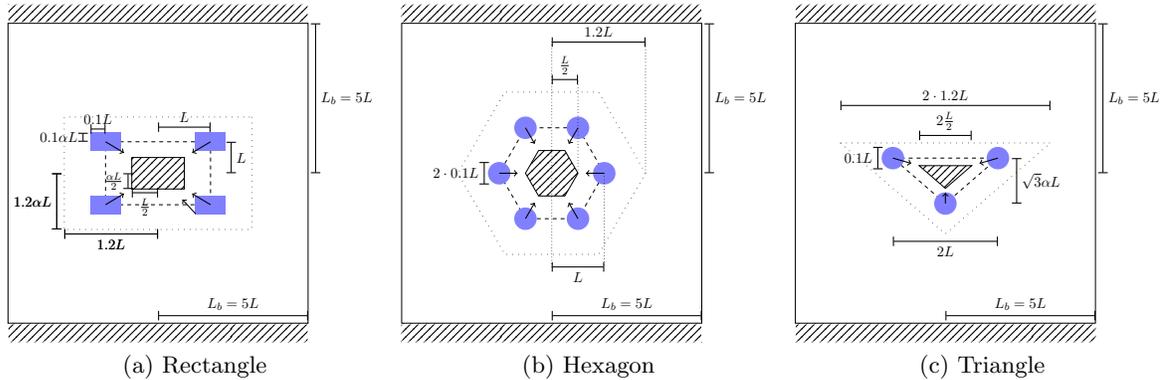
\begin{figure}
    \centering
    \subfloat[][Rectangle]{
        \resizebox{.3\linewidth}{!}{\begin{tikzpicture}
    \newcommand{\figAR}{.6}
    \newcommand{\figPointForceSize}{.4}
    \newcommand{\figInnerSize}{1.4}
    \newcommand{\figJammingSize}{0.7}

    \draw[color=white, pattern=north east lines] ( 4, 4.02) rectangle (-4, 4.5);
    \draw[color=white, pattern=north east lines] ( 4,-4.02) rectangle (-4,-4.5);
    \draw ( 4, 4) rectangle (-4,-4);

    \draw[loosely dotted] ( 2.5, 2.5*\figAR) rectangle (-2.5,-2.5*\figAR) ;
    \draw[dashed] ( \figInnerSize, \figInnerSize*\figAR) rectangle (-\figInnerSize,-\figInnerSize*\figAR) ;

    \foreach \x in {-\figInnerSize,\figInnerSize}
    \foreach \y in {-\figInnerSize*\figAR,\figInnerSize*\figAR}
    {
       \coordinate (c) at (\x, \y);
       \fill[color=blue!50] (\x+\figPointForceSize, \y+\figPointForceSize*\figAR) rectangle (\x-\figPointForceSize, \y-\figPointForceSize*\figAR);
       \draw[->, thick] (c) -- (\x*.65,\y*.65);
    }

    \draw[->, thick] 
    (    \figInnerSize-    \figPointForceSize, -    \figAR*\figInnerSize-    \figAR*\figPointForceSize) -- 
    (.65*\figInnerSize-.65*\figPointForceSize, -.65*\figAR*\figInnerSize-.65*\figAR*\figPointForceSize)    ;
    \draw[pattern=north east lines] (\figJammingSize,\figJammingSize*\figAR) rectangle (-\figJammingSize, -\figJammingSize*\figAR);

    %% Measurements
    %% Simulation cell
    \draw[|-|, thick] (4.2,0) -- node[midway, right] {$L_b=5L$} ++ (0,4);
    \draw[|-|, thick] (0, -3.8) -- node[midway, above] {$L_b=5L$} ++ (4,0);

    %% Activation region
    \draw[|-|, thick] (\figInnerSize+\figPointForceSize+.15,0) -- node[midway, right] {$L$} ++ (0,\figAR*\figInnerSize);
    \draw[|-|, thick] (0, \figAR*\figInnerSize+\figAR*\figPointForceSize+.15) -- node[midway, above] {$L$} ++ (\figInnerSize, 0);

    %% Increased meshing
    \draw[|-|, thick] (0, -2.7*\figAR) -- node[midway, below] {$1.2L$} ++ (-2.5,0);
    \draw[|-|, thick] (-2.7, 0) -- node[midway, left] {$1.2\alpha L$} ++ (0,-2.5*\figAR);
    
    %% Increased meshing
    \draw[|-|, thick] (0, -2.7*\figAR) -- node[midway, below] {$1.2L$} ++ (-2.5,0);
    \draw[|-|, thick] (-2.7, 0) -- node[midway, left] {$1.2\alpha L$} ++ (0,-2.5*\figAR);

    %% Activation region
    \draw[|-|, thick] (-\figInnerSize-\figPointForceSize-.2, \figAR*\figInnerSize) -- node[midway, left] {$0.1\alpha L$} ++ (0,\figAR*\figPointForceSize);
    \draw[|-|, thick] (-\figInnerSize, \figAR*\figInnerSize+\figAR*\figPointForceSize+.1) -- node[midway, above] {$0.1 L$} ++ (-\figPointForceSize, 0);

    %% Jamming region
    \draw[|-|, thick] (-\figJammingSize-.1,0) -- node[midway, left] {$\frac{\alpha L}{2}$} ++ (0,-\figAR*\figJammingSize);
    \draw[|-|, thick] (0, -\figAR*\figJammingSize-.1) -- node[midway, below] {$\frac{L}{2}$} ++ (-\figJammingSize, 0);

\end{tikzpicture}}
        \label{fig:si:geometry:rectangle}
    }
    \subfloat[][Hexagon]{
        \resizebox{.3\linewidth}{!}{\begin{tikzpicture}
    \newcommand{\figAR}{.6}
    \newcommand{\figPointForceSize}{.3}
    \newcommand{\figOuterSize}{2.5}
    \newcommand{\figInnerSize}{1.4}
    \newcommand{\figJammingSize}{0.7}

    \draw[color=white, pattern=north east lines] ( 4, 4.02) rectangle (-4, 4.5);
    \draw[color=white, pattern=north east lines] ( 4,-4.02) rectangle (-4,-4.5);
    \draw ( 4, 4) rectangle (-4,-4);

    \draw[loosely dotted] (0:\figOuterSize) \foreach \x in {60,120,...,360} {  -- (\x:\figOuterSize) };
    \draw[dashed] (0:\figInnerSize) \foreach \x in {60,120,...,360} {  -- (\x:\figInnerSize) };
    \draw[pattern=north east lines] (0:\figJammingSize) \foreach \x in {60,120,...,360} {  -- (\x:\figJammingSize) };

    \foreach \x in {60,120,...,360} {
        \fill[color=blue!50] (\x:\figInnerSize) circle (\figPointForceSize);
        \draw[->, thick] (\x:\figInnerSize) -- (\x:\figInnerSize*.65);
    }

    %% Measurements
    %% Simulation cell
    \draw[|-|, thick] (4.2,0) -- node[midway, right] {$L_b=5L$} ++ (0,4);
    \draw[|-|, thick] (0, -3.8) -- node[midway, above] {$L_b=5L$} ++ (4,0);

    %% Helper line
    \draw[dotted, thin] (0,-2.5) -- (0,3.5);

    %% Activation region
    \draw[dotted, thin] (\figInnerSize,0) -- (\figInnerSize,-2.5);
    \draw[|-|, thick] (0,-2.5) -- node[midway, below] {$L$} (\figInnerSize,-2.5);

    %% Increased meshing
    \draw[dotted, thin] (\figOuterSize,0) -- (\figOuterSize, 3.5);
    \draw[|-|, thick] (0, 3.5) -- node[midway, above] {$1.2L$} (\figOuterSize, 3.5);

    %% Jamming region
    \draw[dotted, thin] (\figJammingSize,0) -- (\figJammingSize, 2.5);
    \draw[|-|, thick] (0, 2.5) -- node[midway, above] {$\frac{L}{2}$} (\figJammingSize, 2.5);

    %% Force field
    \draw[|-|, thick] (-\figInnerSize-\figPointForceSize-.1, -\figPointForceSize) -- node[midway, left] {$2\cdot 0.1L$} ++ (0, 2*\figPointForceSize);

\end{tikzpicture}}
        \label{fig:si:geometry:hexagon}
    }
    \subfloat[][Triangle]{
        \resizebox{.3\linewidth}{!}{\begin{tikzpicture}
    \newcommand{\figAR}{.5}
    \newcommand{\sqrtThree}{1.73205080757}

    \newcommand{\figPointForceSize}{.3}
    \newcommand{\figOuterSize}{2.8}
    \newcommand{\figInnerSize}{1.4}
    \newcommand{\figJammingSize}{0.7}

    \newcommand{\xa}{-1} \newcommand{\ya}{   \figAR*\sqrtThree/3}
    \newcommand{\xb}{ 1} \newcommand{\yb}{   \figAR*\sqrtThree/3}
    \newcommand{\xc}{ 0} \newcommand{\yc}{-2*\figAR*\sqrtThree/3}

    \def\isosceles[#1](#2)%
    {\draw[#1] (#2*\xa, #2*\ya) -- (#2*\xb, #2*\yb) -- (#2*\xc, #2*\yc) -- cycle;}

    \draw[color=white, pattern=north east lines] ( 4, 4.02) rectangle (-4, 4.5);
    \draw[color=white, pattern=north east lines] ( 4,-4.02) rectangle (-4,-4.5);
    \draw ( 4, 4) rectangle (-4,-4);

    \isosceles[dashed](\figInnerSize)
    \isosceles[loosely dotted](\figOuterSize)
    \isosceles[pattern=north east lines](\figJammingSize)

    \fill[color=blue!50] (\figInnerSize*\xa, \figInnerSize*\ya) circle (\figPointForceSize);
    \fill[color=blue!50] (\figInnerSize*\xb, \figInnerSize*\yb) circle (\figPointForceSize);
    \fill[color=blue!50] (\figInnerSize*\xc, \figInnerSize*\yc) circle (\figPointForceSize);

    \draw[->, thick] (\figInnerSize*\xa, \figInnerSize*\ya) -- (.65*\figInnerSize*\xa, .65*\figInnerSize*\ya) ;
    \draw[->, thick] (\figInnerSize*\xb, \figInnerSize*\yb) -- (.65*\figInnerSize*\xb, .65*\figInnerSize*\yb) ;
    \draw[->, thick] (\figInnerSize*\xc, \figInnerSize*\yc) -- (.65*\figInnerSize*\xc, .65*\figInnerSize*\yc) ;

    %% Measurements
    %% Simulation cell
    \draw[|-|, thick] (4.2,0) -- node[midway, right] {$L_b=5L$} ++ (0,4);
    \draw[|-|, thick] (0, -3.8) -- node[midway, above] {$L_b=5L$} ++ (4,0);

    %% Horizontal
    \draw[|-|, thick] (-\figJammingSize, \figOuterSize*\ya+.2) -- node[midway, above] {$2 \frac{L}{2}$} ++ (2*\figJammingSize,0);
    \draw[|-|, thick] (-\figOuterSize, \figOuterSize*\ya+1.0) -- node[midway, above] {$2\cdot 1.2L$} ++ (2*\figOuterSize,0);
    \draw[|-|, thick] (-\figInnerSize, \figOuterSize*\yc-.2) -- node[midway, below] {$2L$} ++ (2*\figInnerSize,0);

    %% Vertical
    %% Activated region
    \draw[|-|, thick] (\figInnerSize*\xb+\figPointForceSize+.2, \figInnerSize*\ya) -- node[midway, right] {$\sqrt{3}\alpha L$} ++ (0,-\figInnerSize*\figAR*\sqrtThree);

    %%% Force field
    \draw[|-|, thick] (-\figInnerSize*\xb-\figPointForceSize-.1, \figInnerSize*\ya-\figPointForceSize) -- node[midway, left] {$0.1L$} ++ (0,2*\figPointForceSize);

\end{tikzpicture}}
        \label{fig:si:geometry:triangle}
    }
    \caption{\textbf{Overview figure of the different geometries used in the simulation.} It should be noted that the figures are not to scale. In these figures, the shaded regions denote areas with no-slip boundaries, and the blue regions mark regions with non-vanishing force density. The outer solid line demarcates the simulation cell, where the left and right boundary have open boundary conditions. The dashed line marks the activation area (compared to the experiment, with no influence on the simulation) and the dotted line indicates the change in meshing from a fine mesh in the center to a coarser mesh outside. For the rectangle and the triangle, $\alpha$ denotes the aspect ratio so that $\alpha=1$ denotes the square and equilateral triangle respectively.}
    \label{fig:si:geometry}
\end{figure}

%% Describe problem, i.e. what model are we trying to solve
Due to the small length scales involved, the Reynolds number is small, so that the system can be described by the Stokes equation. For the sake of completeness, the Stokes equation in \cref{eq:stokes-eq} is repeated here: 
\begin{align}
    \mu \nabla^2\vec u-\nabla p+\vec f&=0,  &\nabla \cdot \vec u &=0,%
\end{align}
where $\mu$ is the fluid viscosity, $\vec u$ is the fluid velocity, and $p$ is the pressure. In the following, several geometric objects will be formally introduced, however, a visual representation of the geometry is presented in \cref{fig:si:geometry} to clarify the following description. The corner force hypothesis introduced in the main text is formally written as 
\begin{align}
    \vec f(\vec x)\equiv \sum_i f_0  \delta[\vec x-(x_i, y_i)] \,(-x_i, -y_i) \label{eq:point-force},%
\end{align}
where $i$ enumerates the corners of the activated polygon. The points $(x_i, y_i)$ describe the coordinates of the corners, so that the delta-function encodes the point-force condition. Using a coordinate system which places the center of mass at the origin of the simulation, the vector $(-x_i, - y_i)$ describes that the forces points towards the center of the structure. For example, if the rectangle with aspect ratio $\alpha$ is simulated, then the origin in the simulation is at the center of the rectangle, and the four corners are located at $(\pm L , \pm \alpha L)$.

%% Boundaries
Additional to the Stokes equation describing the bulk, boundary conditions are required to completely define the problem. As described in the main text, there are two sets of boundary conditions, the first describing a no-slip boundary at the central jamming region. In particular it was assumed that $\vec u(\vec x)\equiv 0$ for all $\vec x$ in the central region with half the linear dimension. For example for the rectangle with aspect ratio $\alpha$, the velocity is forced to vanish in the inner rectangle spanned by the four points $(\pm L/2 , \pm \alpha L/2)$. The second set of boundary conditions is enforced along the boundaries of the short axis of the fluid cell. We chose a simulation domain, so that the boundaries along the short axis coincide with the physical boundaries. Hence, the second set of no-slip boundary conditions is $\vec u[(x,y)]\equiv 0$ if $y=\pm L_{b}$, where $L_b$ is the size of the simulation box. It should be noted that there were no additional boundary conditions implemented for the long axis, since in the experiment these boundaries are sufficiently far away from the light-activated area, so that no boundary effect should be expected. 

%% Give technical details.
% Determine L, determine boundary box, determine grid level
In our particular case, non-dimensionalized values were used with $\mu = f_0= L=1$. In order to have consistent simulation cell size, $L_b = \SI{3}{\milli\meter}/\SI{600}{\micro\meter}\times L=5 L$. Since FEM simulations cannot handle point-forces consistently, the point force was spread out over a small region in the corners. In particular, for the rectangular case, the delta-function in \cref{eq:point-force} was replaced with
\begin{align}
    \delta[(x,y)-(x_i, y_i)]\mapsto\chi\left[\left(|x-x_i|<\frac{L}{10}\right) \land \left(|y-y_i|<\frac{\alpha L}{10}\right)\right],
\end{align}
where $\chi$ is the indicator function, being $1$ where the argument is true, and being $0$ otherwise. This describes small rectangles in the corners describing the non-vanishing forces. For the hexagonal and triangular cases, rather than repeating the same structure, small circles with radius $r=\frac{L}{10}$ were chosen. Formally, this equates to the replacement
\begin{align}
    \delta[(x,y)-(x_i, y_i)]\mapsto\chi\left[\|\vec x - (x_i, y_i)\|<\frac{L}{10}\right].
\end{align}
The meshing contained initially $60\times 60$ equal-distance points, and in a central region of $1.2\times L$ the meshing was refined twice, increasing the local meshing density by a factor of $4$. The Stokes equation was then solved with Taylor-Hoods finite-elements.

%% Implementation: Mention that code is openly available and give github link, reference the FeniCS code once more, and give the tutorial link.
Our implementation of this code can be found on GitHub~\footnote{\url{https://github.com/domischi/StokesFEM}}, which is based on the FeniCS library~\cite{AlnaesBlechta2015a}, and their tutorial on the Stokes equation available at~\footnote{Retrieved August 31, 2020, \url{https://fenicsproject.org/docs/dolfin/2019.1.0/python/demos/stokes-iterative/demo_stokes-iterative.py.html}}. 

%\subsection*{TODO notes}

%\bibliographystyle{aiaa}
%\bibliographystyle{apalike}
%\bibliography{reference,pnas-sample}

\end{document}